\documentclass[aps,pra,twocolumn,groupedaddress]{revtex4-1}
\usepackage[dvips]{graphicx}
\usepackage{amsmath}
\usepackage{amsfonts}

\begin{document}
\renewcommand*{\figurename}{FIG.}
\newtheorem{thm}{Theorem}
\newtheorem{prop}{Proposition}
\newenvironment{proof}[1][\textit{Proof}]{\begin{trivlist}
\item[\hskip \labelsep {#1}]}{\end{trivlist}}

\title{Barycentric measure of quantum entanglement}

\author{Wojciech Ganczarek$^{1}$, Marek Ku{\'s}$^2$, Karol {\.Z}yczkowski$^{1,2}$}

\affiliation{ $^1$Institute of Physics, Jagiellonian University, ul.\
Reymonta 4, 30-059 Krak\'ow, Poland
 \\$^2$Center for Theoretical Physics, Polish Academy of Sciences,
al.\ Lotnik\'ow 32/46, 02-668 Warszawa, Poland}

\date{December 13, 2011}

\begin{abstract}
Majorana representation of quantum states by a constellation of $n$ 'stars' (points on the sphere) can be used to describe any pure state of a simple system of dimension $n+1$ or a permutation symmetric pure state of a composite system consisting of $n$ qubits. We analyze the variance of the distribution of the stars, which can serve as a measure of the degree of non-coherence for simple systems, or an entanglement measure for composed systems. Dynamics of the Majorana points induced by a unitary dynamics of the pure state is investigated.
\end{abstract}


\maketitle

\section{Introduction}

Non--classical correlations between composite quantum systems became a
subject of an intense current research. A particular kind of such
correlations, called {\sl quantum entanglement} attracts special attention of
theoretical and experimental physicists -- see \cite{HHHH09} and references
therein. Entanglement in quantum systems consisting of two subsystems is
nowdays relatively well understood, but several significant questions
concerning multiparticle systems remain still open.

One of the key issues is to workout a practical entanglement measure, used to
quantify the quantum resources of a given state.
 Although  various measures of quantum entanglement are known \cite{MCKB05,BZ06,PV07,HHHH09}, they are usually difficult to compute.
An important class of entanglement measures can be formulated within the geometric approach to the problem \cite{KZ01,BZ06}.
For a given quantum state one can study its minimal distance to the set of separable states. Various distances \cite{BZ06}
can be used for this purpose and the minimization can be performed with respect to the mixed \cite{VP98} or pure \cite{BH01,WG03} separable states.

Working with pure quantum states of a $n$--qubit system it is possible to
distinguish a class of states symmetric with respect to permutations of all
subsystems. This class of symmetric pure quantum states can be identified
with the set of all states of a single system described in a $N=n+1$
dimensional Hilbert space.

Analyzing the space of pure states belonging to $N$ dimensional Hilbert space
it is useful to distinguish the class of {\sl spin coherent} states. These
states $|\theta,\phi\rangle$, labeled by a point on the sphere can be
obtained by action of the Wigner rotation matrix $R_{\theta,\phi}$ on the
maximal weight state $|j,j\rangle$, so they are also called $SU(2)$ coherent
states. Here $j=n/2$ is the maximal eigenvalue of the component $J_z$ of the
angular momentum operator. Making use of the stereographic projection one can
map the sphere into the plane. The coherent states are then labeled by a
complex number $\alpha$ and their expansion in the eigenbasis of $J_z$ reads
\cite{Ag81,ZFG90}
\begin{equation}
|\alpha\rangle = \frac{1}{(1+|\alpha|^2)^{n/2}}
\sum_{k=0}^n {n \choose k}  \alpha^k |k\rangle .
\label{spin} \end{equation}

Consider now an arbitrary state of an $N$--level system,
$|\psi\rangle  =\sum_{k=0}^n c_k |k\rangle $. It can be expanded in
the coherent states representation,
 \begin{equation}
  Q_{\psi}(\alpha)= |\langle \psi |\alpha \rangle |^2
         = \frac{|c_n|^2}{(1+|\alpha|^2)^{n}}
\prod_{i=1}^n |\alpha- z_i|^2
\label{husimi}
\end{equation}
The function $Q_{\psi}(\alpha)$ is called the {\sl Husimi function} (or
$Q$--function) of the state $|\psi\rangle$ and it can be interpreted as a
probability density on the plane. As it can be associated with a polynomial
of order $n$ of a complex argument, it is uniquely represented by the set if
its $n$ roots $z_i$, $i=1,\dots n$, which may be degenerated. With help of
the inverse stereographic projection one can map these points back on the
sphere. The collection of these $n$ points on the sphere, corresponding to
the zeros of the Husimi function (\ref{husimi}), represents uniquely the
state $|\psi\rangle \in {\cal H}_{n+1}$. This approach of  Majorana
\cite{mjr} and Penrose \cite{Pe89} leads to the so-called {\sl stellar
representation} of a state, as each {\sl Majorana point} on the sphere
corresponding to $z_i$, can be interpreted as a star on the sky.

For any coherent state $|\alpha\rangle=|\theta,\phi\rangle$ all $n$ stars sit
in a single point antipodal to the vector pointing in the direction
$(\theta,\phi)$. If the Majorana points are located in a neighborhood of a
single point, the corresponding state is close to be coherent. In contrast, a
generic random state, for which the distribution of stars is uniform on the
sphere \cite{Leb91,BBL92} is far from being coherent. The degree of
non-coherence of $|\psi\rangle$ can be characterized e.g. by its Fubini-Study
distance to the closest coherent state, or by the Monge distance equal to the
total geodesic distance on the sphere all stars have to travel to meet in a
single point \cite{ZS01}.

The stellar representation, originally used for states of a simple system of
$n+1$ levels, can be also used to analyze composite systems consisting of $n$ qubits,
under the assumption that the states are symmetric with respect to permutations
of all subsystems \cite{hay,bast,MGP10,aul,Kol10,RM11}.
Due to this symmetry  the investigation of such states becomes easier
and estimation of their geometric measure of entanglement can be simplified \cite{HK+09}.

Note that any separable state of the $n$--qubit system can be considered as
a coherent state with respect to the composed group $[SU(2)]^n$.
Thus entangled states of a composite system correspond to non-coherent states of the simple system with $n+1$ levels \cite{MZ04},
while the degree of entanglement can be identified with the degree of non-coherence.

In this work we propose to characterize pure quantum states by the position of the barycenter of its Majorana representation.
We will then describe non-classical properties of a state by the average and the variance of the corresponding distribution of the Majorana points.
 The latter quantity has a simple geometric interpretation as a function of
the radius of the barycenter of the stars, which is situated inside the Bloch ball.
The same quantity can thus  be applied to characterize
the degree of non-coherence of a pure state of an $(n+1)$--level system, or simultaneously, as a degree of entanglement
for the corresponding permutation symmetric pure state of $n$--qubit system.

In a similar way the measure of quantumness of a state of $n+1$  dimensional system introduced by Giraud et al. \cite{GBB10} can be used to
characterize the entanglement of permutationally symmetric pure states of $n$ qubits.
 We are also going to study time evolution of
quantum states and an associated dynamics of stars.
In particular we investigate how the barycentric measure of entanglement varies with time.

This work is organized as follows. In section II we review the Majorana representation used for symmetric pure states of a multiqubit
quantum system. In Section III the position of the barycenter of Majorana points and their variance is investigated,
as it may serve as a characterization of quantum entanglement. A family of $n$-qubit states
maximally entangled with respect to this measure is studied in section IV.
Unitary dynamics of quantum states and the associated dynamics of Majorana points on the sphere is analyzed in Section V.

\section{Majorana representation of multi--qubit states}

Consider a $n$-qubit pure state $|\psi\rangle$
symmetric with respect to permutation of subsystems:
\begin{equation}
 \mid\psi\rangle=\frac{1}{\sqrt{K}}\sum\limits_\pi\mid\phi_{\pi(1)}
\rangle\mid\phi_{\pi(2)}\rangle\cdots\mid\phi_{\pi(n)}\rangle.
\label{eq:maj}
\end{equation}
The sum is taken over all permutations $\pi$, while the normalization
constant reads
$$K=n!\; \sum\limits_\pi\prod\limits_{i=1}^n\mid\langle
\phi_i\mid \phi_{\pi(i)}\rangle\mid.$$ Any one-qubit pure state can be
represented as $\mid\phi_{i}\rangle=\cos\frac{\theta_i}{2}\mid0\rangle+
e^{i\Phi_i}\sin\frac{\theta_i}{2}\mid1\rangle$.
Making us of the the Dicke states
$\mid S_{n,k}\rangle$ with $k=0,1,\dots, n$  \cite{DK54},
\begin{equation}
\mid S_{n,k}\rangle= {n \choose k}^{-\frac{1}{2}}\sum\limits_\rho
P_\rho \mid0\rangle^{\otimes n-k}\mid1\rangle^{\otimes k},
\label{eq:dick}
\end{equation}
we can represent the state $|\psi\rangle$ as their superposition,
\begin{equation}
| \psi\rangle=\sum_{k=0}^nd_k  | S_{n,k}
\rangle=\sum\limits^n_{k=0}d_k {n \choose
k}^{-\frac{1}{2}} \! \sum\limits_{\rho} P_\rho |
 0\rangle^{\otimes n-k} |1\rangle^{\otimes k}.
\label{eq:dick2}
\end{equation}

Here $d_k$ denote complex coefficients, $n \choose k$ is the binomial
coefficient and the sum $\sum\limits_\rho P_\rho$ goes over all states of
$n$-qubits with exactly $k$ qubits in the state $|1\rangle$ and $n-k$ qubits
in the state $|0\rangle$. It comes out that quotients of the coefficients
 $\frac{\cos\frac{\theta_i}{2}}{e^{i\Phi_i}\sin\frac{\theta_i}{2}}=e^{-i\Phi_i}  \cot  (\theta_i /2)$ determining
 the one--qubit state  $|\phi_i\rangle$ can be obtained as roots $z_i$ of the polynomial
\begin{equation}
  P(z)=\sum\limits_{k=1}^nd_k(-1)^k{n \choose k}^{\frac{1}{2}}z^k.
\label{eq:poly}
\end{equation}

The polynomial defined above has degree $D\leq n$ and consequently as many
roots.In case of $D<n$ the remaining $(n-D)$ coefficients $\cos\frac{\theta_i}{2}$ we set to unity \cite{bast}.

 In this way we can represent any n-qubit state symmetric with respect to
 permutations as $n$ points on the Bloch sphere related to spin-$\frac{1}{2}$
 (1-qubit) states $\mid\phi_{i}\rangle=\cos\frac{\theta_i}{2}\mid0\rangle+
 e^{i\Phi_i}\sin\frac{\theta_i}{2}\mid1\rangle$.
These $n$ points on the
 Bloch sphere define a {\sl stellar representation} of a state and are called  stars or \textit{Majorana points} (MP), while  polynomial (\ref{eq:poly})
 is called the \textit{Majorana polynomial} \cite{mjr}.

Note that the same constellation of $n$ stars determines on one hand a pure state of the simple system described in $N=n+1$ dimensional
 Hilbert space \cite{Ag81,Pe89,BZ06}. On the other hand, it describes also a permutationally symmetric pure state of a $n$ qubit system \cite{hay,MGP10,aul} and
belongs to the Hilbert space of dimension $2^n$.

\begin{center}
\begin{figure}
\hskip -0.4cm
\includegraphics[height=5cm]{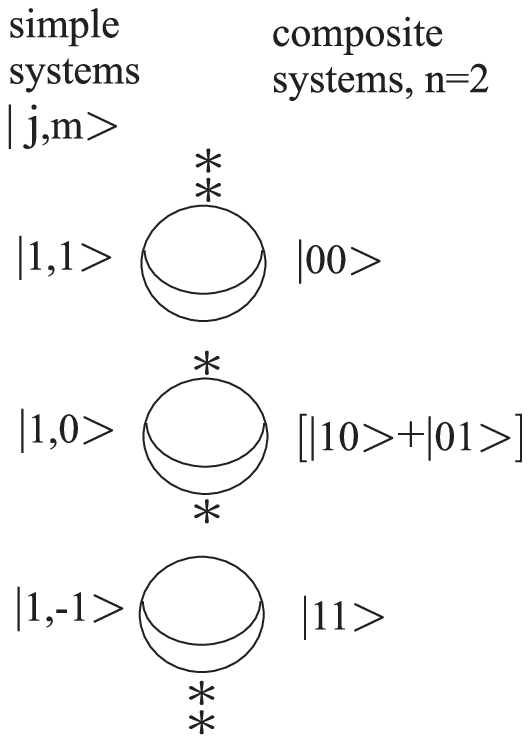}\hskip 0.3cm
\includegraphics[height=5.7cm]{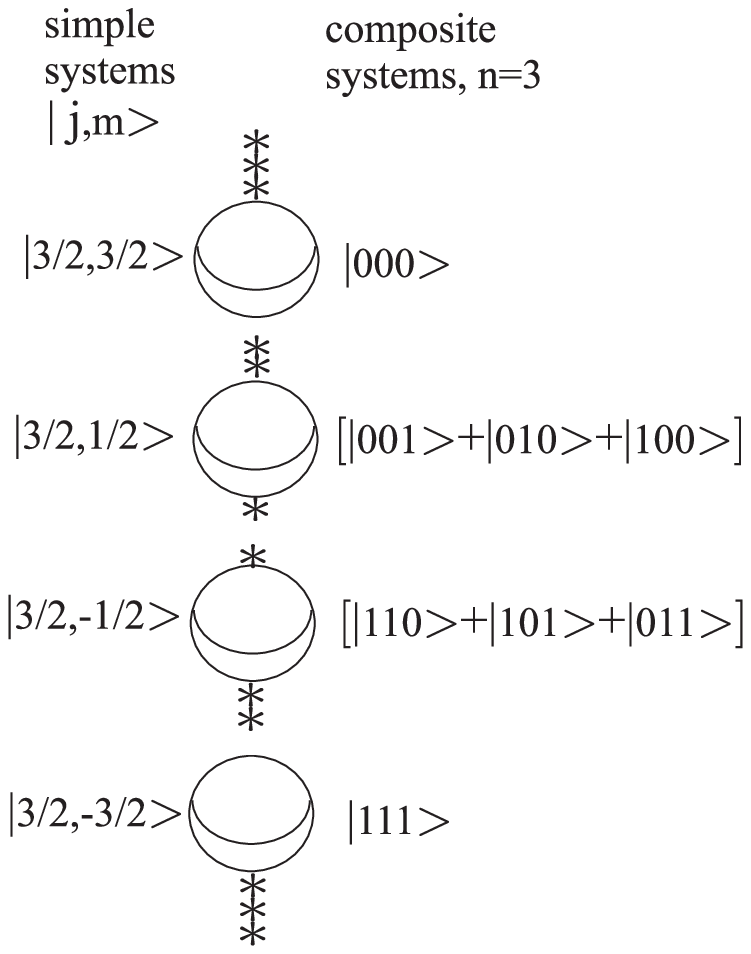}
\caption{Stellar representation of the orthogonal basis
in  $(n+1)$ dimensional Hilbert space describing a simple system
(states $|j,m\rangle$ with $j=n/2$)
with the corresponding basis in the
subspace of permutation symmetric states of $n$--qubit systems
plotted for $n=2$ and $n=3$.}
\label{fig:stars}
\end{figure}
\end{center}

To establish a one--to--one link between both problems it is sufficient to identify bases in both spaces. In the case of a simple system described in
 the Hilbert space of size $n+1$ we select the standard eigenbasis of the angular momentum operator $J_z$,
\begin{equation}
|j,m\rangle, \ \ m=-j,\dots, j
\label{eq:jmbasis}
\end{equation}
with $j=n/2$. A state from this basis is represented by a constellation of $n/2+m$ stars at the north pole and the remaining $n/2-m$ stars at
the south pole as shown in Fig.~\ref{fig:stars}. Thus the state $|j,m\rangle \in {\cal H}_N$ can be identified with the Dicke
state $|S_{n,k}\rangle \in {\cal H}_2^{\otimes n}$ defined in (\ref{eq:dick}) with $k=n/2-m$. In this way any constellation of stars can be used
 to describe pure quantum state of two different physical systems:
an $(n+1)$-level system and a symmetric state of $n$ qubits.
In following sections we show that this link can be extended also for quantum dynamics.
Any discrete unitary dynamics of an $n$--qubit system,
which preserves the permutation symmetry,  can be represented by an unitary matrix of size $2^n$ with a block structure
in the computational basis. The diagonal block of size $n+1$ is unitary and it
defines the corresponding dynamics of a simple quantum system with $n+1$ levels.

\label{maj}
\section{Barycenter as an entanglement measure}
We propose a quantity designed to characterize entanglement for permutation symmetric states. It is based on the distance between the barycenter of the Majorana points representing a permutation-symmetric state and the center of the Bloch ball.
We will show that this quantity
vanishes on separable states and it does not increase
under local operations which preserve the symmetry.

\subsection{Entanglement measure for permutation symmetric states}
For any permutation symmetric state (\ref{eq:maj}) with
$\mid\phi_{i}\rangle=\cos\frac{\theta_i}{2}\mid0\rangle+
e^{i\Phi_i}\sin\frac{\theta_i}{2}\mid1\rangle$ let us define:
\begin{equation}
 E_B=1-d^2,
\end{equation}
where $d$ is the distance between the barycenter of the Majorana points
representing the state and the center of the Bloch ball:
\begin{equation}
d= \left| \frac{1}{n}\sum\limits_{i=1}^n (\sin\theta_i\cos\Phi_i ,
\,\,\,\sin\theta_i\sin\Phi_i , \,\,\, \cos\theta_i)\right|,
\end{equation}
where $| x |$ denotes the length of a vector $x$.

Let us call $E_B$ the {\sl barycentric measure} of entanglement.
This quantity can be interpreted as the
variance of all Majorana points representing the state.
Indeed, since a vector directed toward any
point at the unit sphere has the unit length we have
$E_B=D^2X=\langle X^2\rangle- \langle X\rangle^2=1-\langle X\rangle^2=1-d^2$.
It is easy to see that $E_B \in [0,1]$.

Note that the quantity $E_B$ can be used to characterize
the symmetric states of the $n$--qubit system or
the states of a single system of size $N=n+1$.
For a separable state $|\psi\rangle=|\phi\rangle^{\otimes n}$
(or an $SU(2)$--coherent state of a single quNit)
all stars are located in a single point, so $d=1$ and $E_B=0$. Moreover,
 a generic random state is likely to be highly entangled (or highly non--coherent)
as the Majorana points are distributed almost uniformly
 at the Bloch sphere \cite{Leb91,BBL92}, which implies $d \approx 0$ and $E_B\approx 1$.

A useful entanglement measure should not increase under local operations.
Since the barycentric measure $E_B$ is defined only for symmetric states so
monotonicity can be checked only for local unitary operations
which preserve the permutation symmetry.
Consider two $n$-qubit permutation symmetric states $|\psi\rangle$,
$|\phi\rangle$ connected by a local operation. This means that there exist
invertible unitary operators $A_i$ such that $|\psi\rangle=A_1\otimes\cdots\otimes
A_n|\phi\rangle$. Mathonet \textit{et al.} proved in \cite{math} that in fact
we may find a single invertible operator $A$ for which
$|\psi\rangle=A^{\otimes n}|\phi\rangle$. Geometrically such an operation
corresponds to a rotation of each Majorana points around the same axis by the
same angle or, equivalently, a rigid rotation of the
whole Bloch sphere. Obviously thus the radius of the barycenter d, and
consequently the barycentric measure $E_B$, do not change. This concludes a
proof of
\begin{prop}
The barycentric measure $E_B$ is invariant under local operations preserving
the permutation symmetry.
\end{prop}

\begin{center}
\begin{figure}
\includegraphics[height=5.5cm]{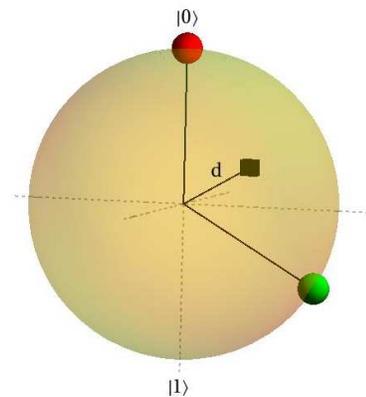}
\caption{(color online) Majorana representation of a symmetric 2-qubit state
of the form $|\phi_1\rangle=|0\rangle$ and
$|\phi_2\rangle=\cos\frac{\theta}{2}|0\rangle+\sin\frac{\theta}{2}|1\rangle$
with $\theta=\frac{2}{3}\pi$. Majorana points
are shown as red and green balls. The black
cube denotes the barycenter of MP located inside the Bloch ball
and its distance to the center of the ball is equal to $d$.}
\label{fig:2q}
\end{figure}
\end{center}
\begin{center}
\begin{figure}
\includegraphics[height=5cm]{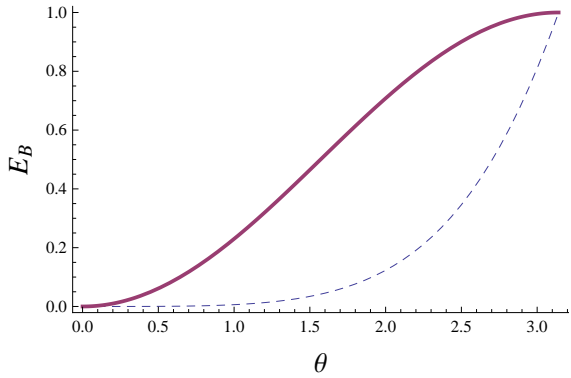}
\caption{Dependence of entanglement measures for a symmetric 2-qubit state
determined by $|\phi_1\rangle=|0\rangle$ and
$|\phi_2\rangle=\cos\frac{\theta}{2}|0\rangle+\sin\frac{\theta}{2}|1\rangle$.
Thick curve denotes the barycentric measure $E_B$,
 while the geometric measure $E_G$ is represented by the dashed curve.}
\label{fig:2q-m}
\end{figure}
\end{center}
\subsection{Examples}
We shall calculate the barycentric measure $E_B$ for certain
exemplary quantum states
and make comparison with the {\sl geometric measure} of entanglement $E_G$
defined as a function of the distance from the analyzed state $\phi$ and the
set of separable states $\mathcal{H}_{SEP}$.
Such a quantity, first proposed by Brody and Hughston \cite{BH01}
and later used in \cite{WG03}, can be defined by
\begin{equation}
E_G(|\phi\rangle)=\min_{|\lambda\rangle\in\mathcal{H}_{SEP}}
\log_2\Big(\frac{1}{|\langle\lambda|\phi\rangle|^2}\Big),
\end{equation}

Let us start with the two-qubit case. Without loss of generality we may set
$|\phi_1\rangle=|0\rangle$ and
$|\phi_2\rangle=\cos\frac{\theta}{2}|0\rangle+\sin\frac{\theta}{2}|1\rangle$.
The special case $\theta=\frac{2}{3}\pi$ is shown in Fig.~\ref{fig:2q}.
Direct calculation gives for this family of states
$E_B=1-\cos^2\frac{\theta}{2}$. Fig.~\ref{fig:2q-m} shows how $E_B$ and
$E_G$ change with the angle $\theta$. We see that in this case $E_B\geq E_G$,
but for a product state ($\theta=0$) and for the maximally entangled state
($\theta=\pi$) both quantities coincide.
\begin{center}
\begin{figure}
\includegraphics[height=5cm]{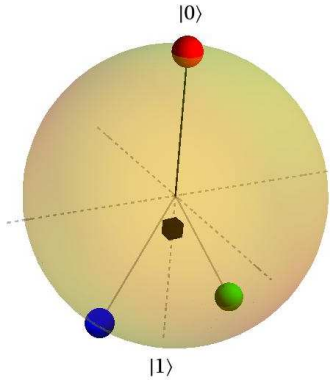}
\caption{(color online) Majorana representation of a symmetric 3-qubit state
$|\phi_1\rangle=|0\rangle$, $|\phi_2\rangle=\cos\frac{\theta}{2}|0\rangle-\sin\frac{\theta}{2}|1\rangle$ and $|\phi_3\rangle=\cos\frac{\theta}{2}|0\rangle+\sin\frac{\theta}{2}|1\rangle$ with $\theta=\frac{5}{6}\pi$. MP are shown as red, green and blue balls. The black cube denotes the barycenter of MP.}
\label{fig:3q}
\end{figure}
\end{center}
%%%
\begin{center}
\begin{figure}
\includegraphics[height=5cm]{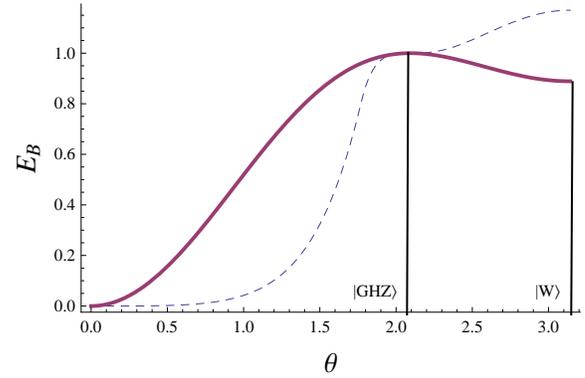}
\caption{Entanglement measures of symmetric 3-qubit state of the form such that $|\phi_1\rangle=0$, $|\phi_2\rangle=\cos\frac{\theta}{2}|0\rangle-\sin\frac{\theta}{2}|1\rangle$ and $|\phi_3\rangle=\cos\frac{\theta}{2}|0\rangle+\sin\frac{\theta}{2}|1\rangle$. Thick line - $E_B$, dashed line - $E_G$. Result for $E_G$ from \cite{aul}.}
\label{fig:3q-m}
\end{figure}
\end{center}

Analysis of $3$--qubit states exhibits a more interesting situation.
Consider a one--parameter family of $3$-qubit states, $|\phi_1\rangle=|0\rangle$,
$|\phi_2\rangle=\cos\frac{\theta}{2}|0\rangle-\sin\frac{\theta}{2}|1\rangle$
and
$|\phi_3\rangle=\cos\frac{\theta}{2}|0\rangle+\sin\frac{\theta}{2}|1\rangle$.
The case $\theta=\frac{5}{6}\pi$ is shown in Fig.~\ref{fig:3q}.
Using
simple geometry we obtain that for such states the barycentric measure reads
$E_B=1-|\frac{2\cos\theta+1}{3}|^2$. Fig.~\ref{fig:3q-m} shows how $E_B$ and
$E_G$ change with the angle $\theta$. Note  a significant difference between $E_B$
and $E_G$. The barycentric measure $E_B$ reaches its maximum at
$\theta=\frac{2}{3}\pi$, which corresponds to the GHZ state and we have
$E_B(|GHZ_3\rangle)=E_G(|GHZ_3\rangle)=1$. However for $\theta>\frac{2}{3}\pi$
$E_G$ increases and reaches its maximal value at $\theta=\pi$, at the
state $|W_3\rangle = \frac{1}{\sqrt{3}}(|001\rangle + |010\rangle + |100\rangle)$.
Moreover, it is easy to see that the state $|GHZ_3\rangle$
is maximally entangled with respect to the barycentric measure
$E_B$: for three points at the sphere their barycenter could be in the center
of the sphere if and only if they form an equilateral triangle. \label{ghz}
\begin{figure}
\includegraphics[width=6cm]{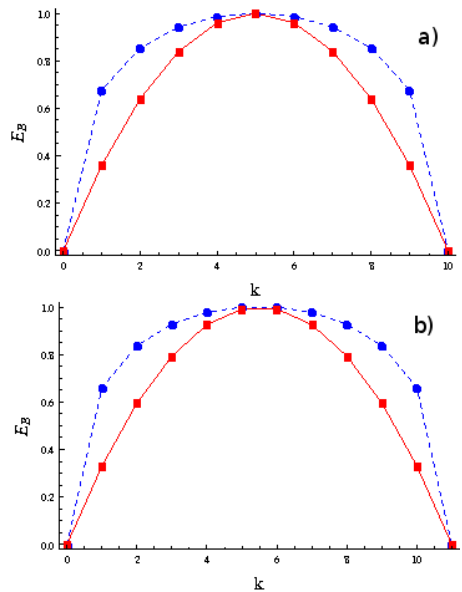}
\caption{(color online) Dependence of entanglement of Dicke states $|\,S_{n,k}\rangle$,
as a function of $k$ for a) $n=10$ and b) $n=11$.
Red squares correspond to the barycentric measure $E_B$,
and blue circles denote the geometric measure $E_G$ normalized to unity
for maximally entangled states - lines are plotted to guide the eye.}
\label{fig:dick}
\end{figure}

As a third example we consider the Dicke states $|\,S_{n,k}\rangle$
(\ref{eq:dick}). They are characterized by $k$ stars at
the south pole and the
remaining $n-k$ Majorana point occupying the north pole
of the Bloch sphere.

The length of the radius of the barycenter is thus easily calculated as
$d=|\frac{n-2k}{n}|$. We would like to make the comparison with $E_G$ in a
slightly different way employing results of Hayashi \textit{et al.} \cite{hay}
who showed that the product state closest to $|\,S_{n,k}\rangle$  reads
\begin{eqnarray}
|\Lambda\rangle&=&\left(\sqrt{\frac{n-k}{n}}|0\rangle+
\sqrt{\frac{k}{n}}|1\rangle\right)^{\otimes n},
\end{eqnarray}
hence
\begin{eqnarray}
E_G(S_{n,k})&=&\log_2\left(\frac{\big(\frac{n}{k}\big)^k
\big(\frac{n}{n-k}\big)^{n-k}}{\binom{n}{k}}\right).
\end{eqnarray}
The projection on z axis of $\Lambda$ has the
length equal $|\frac{n-2k}{n}|$, just like the length of the radius $d$ (see above).
For an even $n$ the state $|\,S_{n,n/2}\rangle$ is maximally entangled with
respect to the measure $E_G$ among all Dicke states; for odd $n$ there are
two such states, $|\,S_{n,(n-1)/2}\rangle$ and $|\,S_{n,(n+1)/2}\rangle$.
The same states are extremal with respect to the barycentric measure.

 Moreover, for an even $n$ the state
$|\,S_{n,n/2}\rangle$ belongs to the class of the maximally entangled
$n$-qubit states as its barycentric measure $E_B$ is equal to unity.
A comparison between the measures $E_B$ and $E_G$ is shown in Fig.~\ref{fig:dick}
for Dicke states with $n=10$ and $n=11$. By construction
the same results hold for the states of the standard $|j,m\rangle$ basis
in the space of $N=n+1$ level systems, with $j=n/2$ and $m=k-n/2$.

\section{Family of maximally entangled states}  
It is interesting to look for states which are maximally entangled with respect to the barycentric measure $E_B$. The answer is simple for a small number of qubits. For $n=2$ a constellation of two stars on the sphere has its barycenter in the center of the sphere if and only if it consists two antipodal points on the sphere. Therefore the class of maximally entangled $2$-qubit states with  $E_B=1$ coincides with all states equivalent to the Bell state up to a local unitary transformation.

\subsection{Two and three qubits}

For the $2$--qubit problem one can establish the following property. \begin{prop} There is no universal unitary operator $U(\theta)$ which transforms any separable state with both MP at $\theta_0$ into a state with MPs at $\theta_0\pm\theta$ for a given angle $\theta$.
\end{prop}
Indeed, assume that $A$ is such an operator. Then, in particular, it would
transform the states $|00\rangle$ and $|11\rangle$ into the Bell state
$|\Psi_+\rangle=\frac{1}{\sqrt{2}}(|01\rangle+|10\rangle)$, which
is impossible since a unitary operator cannot transform two orthogonal
states into the same state.

\begin{figure}
\includegraphics[width=5.0cm]{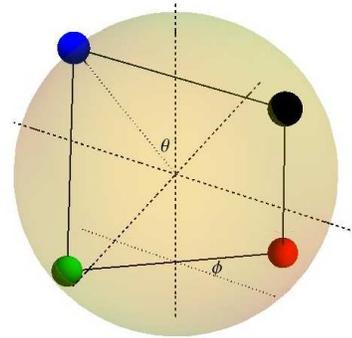}
\caption{Parameterization of the family of the maximally entangled $4$-qubit
states. The dashed lines are the $3$ axes of the coordinate system. The full
lines are the $4$ sides of the rectangle. The upper side is parallel to the
$X$-axis, the bottom side is rotated by an angle $\Phi$. The dotted lines are
used to mark the angles: the bottom dotted line is parallel to the $X$-axis
i.e., at the position of the bottom side of rectangle without rotation.}
\label{fig:oz}
\end{figure}

The $3$-qubit case was discussed in the preceding section \ref{ghz} with the conclusion that the class of maximally entangled $3$-qubit states consist of the GHZ state and the states unitarily equivalent. The case of $4$-qubit states reduces to a geometric problem of placing $4$ points on the sphere in such a way that their barycenter is located in the center of the sphere. Consider a rectangle inscribed into a sphere, belonging to the XZ plane, so that its barycenter is the center of the sphere. Without changing the position of the barycenter we can vary  the lengths of its sides by changing the angle $\theta$ between one side of the rectangle and its diagonal (see Fig. \ref{fig:oz}). We can also rotate one of its sides (the bottom side in Fig.~\ref{fig:oz}) varying the angle $\Phi\in[0,\pi]$.

\subsection{Four qubits}

Assume now that these four points represent a certain $4$-qubit permutation symmetric state $|\psi_{rec}\rangle$. It means that $|\psi_{rec}\rangle$ is the symmetrization (\ref{eq:maj}) of the tensor product of
\begin{eqnarray}\nonumber
|\phi_1\rangle&=&\cos\frac{\theta}{2}|0\rangle+e^{i\Phi}\sin\frac{\theta}{2}|1\rangle, \\ \nonumber
|\phi_2\rangle&=&\cos\frac{\theta}{2}|0\rangle-e^{i\Phi}\sin\frac{\theta}{2}|1\rangle, \\
|\phi_3\rangle&=&\cos\frac{\pi-\theta}{2}|0\rangle+\sin\frac{\pi-\theta}{2}|1\rangle, \\ \nonumber
|\phi_4\rangle&=&\cos\frac{\pi-\theta}{2}|0\rangle-\sin\frac{\pi-\theta}{2}|1\rangle,
\end{eqnarray}
where $\theta\in[0,\pi/2]$, $\Phi\in[0,\pi]$. Consequently,
\begin{eqnarray}\nonumber
&|\psi_{rec}\rangle=
\frac{e^{i\Phi}}{\sqrt{K}} \Big(6e^{-i\Phi}\sin^2\theta|0000\rangle
+6e^{i\Phi}\sin^2\theta|1111\rangle+&\\
&\big[4i\cos\theta \sin\Phi-2(\cos^2\theta+1)\cos\Phi\big]
\sum\limits_\pi|0011\rangle\Big),&
\label{max4}
\end{eqnarray}
where $K$ is the normalization factor. The state $|\psi_{rec}\rangle$ defined above and all locally equivalent states form a class of $4$-qubit states maximally entangled with respect to the barycentric measure $E_B$.
This class contains $|W_4\rangle$ state 
(which we identify with $|S_{4,2}\rangle$), obtained for $\theta=\Phi=0$,
such that all four Majorana points belong to a single line:
two stars are localized at the north pole and two other
at the south pole.
The choice  $\theta=\pi/2$ and $\Phi=\frac{\pi}{2}$
produces $|GHZ_4\rangle=\frac{1}{\sqrt{2}} [|0000\rangle + |1111\rangle ]$
represented by stars in four corners  of a square at the equator.
Setting $\theta=\pi/4$ and $\Phi=0$ we obtain a state locally equivalent to
$|GHZ_4\rangle$ with stars situated on a plane perpendicular to the equatiorial one.
The tetrahedron state $|\psi_{\rm tetr}\rangle$ corresponds
to  $\theta=\arccos (1/\sqrt{3})$ and $\Phi=\frac{\pi}{2}$.

\begin{figure}
\includegraphics[width=5.6cm]{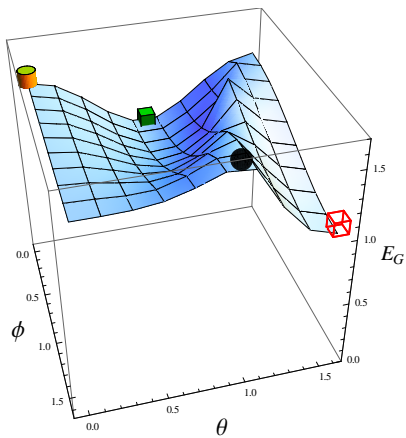}	
\caption{(color online) Geometric measure of entanglement for the family (\ref{max4})
of four--qubit states for which $E_B=1$
parameterized by angles $(\theta, \Phi)$ is not smaller than $1$.
Cylinder represents  the state $|W_4\rangle$, a ball denotes the tetrahedron
state for which $E_G$ is maximall,
while two cubes represent two equivalent $|GHZ_4\rangle$ states (empty cube - $|GHZ_4\rangle$ with stars situated on the equatorial plane, full cube - rotated $|GHZ_4\rangle$, situated on a plane perpendicular to the equatiorial one).}
\label{fig:eceb}
\end{figure}

To compare both measures of entanglement, following \cite{aul} we computed numerically
the geometric measure of entanglement $E_G$ for the entire family (\ref{max4})
of extremal states, for which $E_B=1$ (see Fig.~\ref{fig:eceb}). All states from this class
are characterized
by $E_G\ge 1$, and the minimum is attained for a $|GHZ_4\rangle$ state, for which
 $E_G=1$. For a $|W_4\rangle$ state one has $E_G(|W_4\rangle)=\log_2(8/3)\approx 1.415$ \cite{hay},
while the maximum is obtained for the tetrahedron state
and  $E_G(|\psi_{\rm tetr}\rangle)=\log_2 3\approx 1.585$ \cite{aul}.
This observation suggests that the both measures are well correlated,
so the barycentric measure of entanglement, simple to evaluate,
could be used to characterize the state in parallel to the
geometric measure.

Note that the states
called 'Queen of Quantum', the most distant from the set of product states with respect to the Hilbert-Schmidt \cite{GBB10} or Bures \cite{MGP10} distance,
are  for small values of $N$ described by symmetric constellations of the Majorana points,  so they are maximally entangled with respect to the barycentric measure $E_B$.

\subsection{Composition of states}

Consider two pure states of a simple, $N$--level system,
$|{\psi}_1\rangle $ and $|{\psi}_2\rangle $
in ${\cal H}_N$. A {\sl composition}
 of these states, $|\psi_1\rangle \odot |\psi_2\rangle $,
defined in \cite{BZ06}, forms a pure state belonging to a $2N-1$ symmetric
subspace of the tensored Hilbert space, ${\cal H}_N \otimes {\cal H}_N$.
Both initial states are described by $n=N-1$ stars,
and their composition, represented by $2n$ stars,
is defined by adding the stars together. In the similar way one can
compose the states from two Hilbert spaces of different dimensions.
More formally, one can define the composition by its Husimi
distribution $Q(\alpha)$,
 which is given by the product of two Husimi distributions
representing individual states,
\begin{equation}
 Q_{|\psi_1\rangle \odot |\psi_2\rangle } \propto
Q_{|\psi_1\rangle }Q_{|\psi_2\rangle } \ .
\label{comphus}
 \end{equation}

The same idea of composition of two pure states of a simple system
described by the stellar representation can be now used
for the symmetric states of several qubits.
Consider a permutation symmetric $n$-qubit state
 $|{\psi} \rangle \in {\cal H}_2^{\otimes n}$ and
another $m$-qubit state
 $|{\phi} \rangle \in {\cal H}_2^{\otimes m}$.
Their composition, written $|\psi\rangle \odot |\phi\rangle
\in {\cal H}_2^{\otimes n+m}$, represents a state of $n+m$ qubits,
and is defined by the entire sum of $n+m$ Majorana points on the sphere.

Observe that this notion allows us to write the symmetric Bell state,
$|\Psi_+\rangle=\frac{1}{\sqrt{2}}(|01\rangle+|10\rangle)= |1\rangle \odot
|0\rangle$. Introducing two, one--qubit superposition states, $| \pm
\rangle=\frac{1}{\sqrt{2}}(|0\rangle \pm |1\rangle)$ we see that another Bell
state $|\Phi_-\rangle=\frac{1}{\sqrt{2}}(|00\rangle-|11\rangle)$ is
equivalent to the composition $|+\rangle \odot |-\rangle$. On the other hand,
any symmetric separable state of $n$--qubit system can be represented by the
composition performed $n-1$ times, $|\phi, \dots , \phi \rangle= |\phi\rangle
\odot \dots \odot  |\phi\rangle$, as in this case all the stars do coincide.

Take any two figures with the same barycenter. Superposing them one obtains
another figure with the same barycenter. This fact implies a useful
\begin{prop}.
Consider two permutation symmetric states of any number of qubits, maximally
entangled with respect to the barycentric measure, $E_B(\phi)=E_B(\psi)=1$.
Then their composition is also maximally entangled, $E_B(\phi \odot \psi)=1$.
\label{prop3}
\end{prop}

To watch this proposition in action consider
the composition of two maximally entangled  Bell states,
$|\Psi_+\rangle$ and  $|\Phi_-\rangle$.
Their composition,
$|\Psi_+\rangle \odot |\Phi_-\rangle = |0\rangle \odot |1\rangle
\odot |+\rangle \odot |-\rangle$ represents
the state formed by a square belonging to the plane
containing the meridian of the sphere. It is then locally equivalent to the
state $|GHZ_4\rangle$, represented by a square inscribed into the equator,
also maximally entangled with respect to the barycentric measure.

The notion of a composition allows us two write
down a generic random state of an $n$ qubits system
\begin{equation}
 |\psi_{\rm rand}\rangle  \ : = \
|\phi_1\rangle \odot |\phi_2\rangle \odot
 \dots \odot  |\phi_n\rangle,
\label{rand1}
 \end{equation}
where $\phi_i$, $i=1,\dots n$,
denotes one--qubit state,
generated by a vector taken randomly with respect to the
uniform measure on the sphere.
For large $n$ such a generic state is highly entangled
with respect to $E_B$, as the barycenter of the points will
be located close to the center of the sphere.

Making use of proposition \ref{prop3}
one may also design a random state,
for which the barycentric measure is equal to unity.
It is sufficient to compose several maximally entangled Bell states
represented by a random collection of pairs
of antipodal points,
\begin{equation}
 |\psi_{\rm rand}'\rangle  \ : = \
|\phi_1\rangle \odot |{\bar \phi_1}\rangle \odot
 \dots \odot
|\phi_{n/2}\rangle \odot |{\bar \phi_{n/2}}\rangle,
\label{rand2}
 \end{equation}
The number $n$ of qubits is assumed to be even
and the directions labeled by
$\phi_i$ and  $\bar{\phi_i}$ are antipodal, so the state
$|\phi_i\rangle \odot |{\bar \phi_i}\rangle$
is maximally entangled and locally equivalent to the Bell state.
Their composition produces  the state  $|\psi_{\rm rand}'\rangle$,
which is characterized by the maximal possible value
of the barycentric measure, $E_B=1$.

\section{Dynamics of Majorana points}
\label{psi1}

\subsection{Interpolation between states}
In this section we analyze the parametric dynamics of the
Majorana points when a state changes with some parameter $\beta$.
A related problem was analyzed by Prosen \cite{Pr96}
who analyzed the parametric dynamics of stars on the complex plane
representing a quantum state under a change of paremeters of the system.
We will allow the pure state to evolve in time, so the parameter
$\beta$ can be interpreted as time.

 As a first example consider the state
$|\psi_1\rangle=\cos\beta
|00\rangle+\frac{\sin\beta}{\sqrt{2}}(|01\rangle+|10\rangle)$ with
$\beta\in[0,\pi]$. It can be written in the form (\ref{eq:maj}) with the help
of the one-qubit states,
\begin{eqnarray}\nonumber
\mid\phi_1\rangle&=&\mid0\rangle  \\
\mid\phi_2\rangle&=&\frac{\sqrt{K(\beta)}}{2}\cos\beta\mid0\rangle+
\sqrt{\frac{K(\beta)}{2}}\sin\beta\mid1\rangle,
\label{eq:e1}
\end{eqnarray}
where $K(\beta)=\frac{4}{2-\cos^2\beta}$. A unitary operator $U$ which generates
such a dynamics is not uniquely determined. If we assume that the operator $U$
transforming $|00\rangle$ into $|\psi_1\rangle$ has the form
$U=\mathbb{I}\otimes M_1+M_2\otimes\mathbb{I}$, we arrive at two
possibilities,
\begin{eqnarray}\nonumber
U_1&=&\cos\beta\,\mathbb{I}\otimes\sigma_z -i\frac{\sin\beta}{\sqrt{2}}
\,\sigma_y\otimes\mathbb{I}+\frac{\sin\beta}{\sqrt{2}}
\,\mathbb{I}\otimes\sigma_x, \\
U_2&=&\cos\beta\,\sigma_z\otimes\mathbb{I}-i\frac{\sin\beta}{\sqrt{2}}
\,\mathbb{I}\otimes\sigma_y+
\frac{\sin\beta}{\sqrt{2}}\,\sigma_x\otimes\mathbb{I}.
\label{eq:u12}
\end{eqnarray}
The operators $U_1$ and $U_2$ are similar: the only
difference is the order of the operators in the tensor product.
Operating on a permutation symmetric state we can act either on the first
or on the second subsystem obtaining the same outcome.

\begin{figure}
\includegraphics[width=6cm]{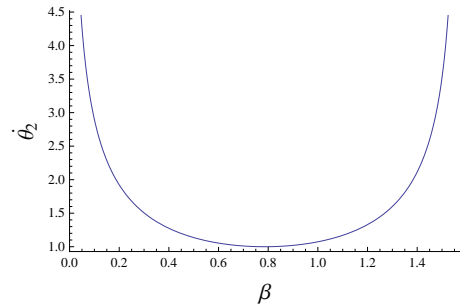}
\caption{Dependence of the velocity $V =\partial_\beta \theta$ of the star
traveling across the sphere  defined as
a function of the phase $\beta$ for the state $|\psi_1\rangle=\cos\beta
|00\rangle+\frac{\sin\beta}{\sqrt{2}}(|01\rangle+|10\rangle)$.
 The minimal $V$ is achieved for a Bell state, $\beta=\pi/2$,
while the maximal velocity is obtained in vicinity of
product states,  $\beta \to 0$ and $\beta \to \pi$.
} \label{fig:ve1}
\end{figure}

\begin{figure}
\includegraphics[width=5cm]{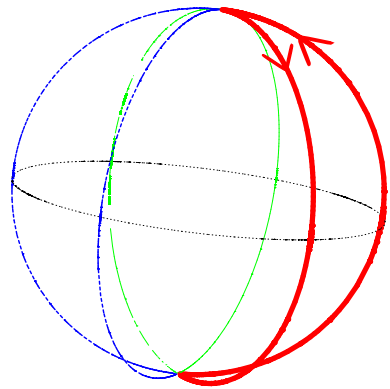}
\caption{(color online) Trajectories of Majorana points for the state
$\frac{1}{\sqrt{K}}\big(\cos\beta\mid000\rangle+\sin\beta\mid
GHZ\rangle\big)$. Regular green, dashed blue and thick red lines represent trajectories of 3 MPs
for $\beta\in[0,\pi]$, black dashed line represents the equator.
Three stars travel from the north pole to the south pole
and at the equator, for $\beta=\pi/2$, they form a $|GHZ\rangle$ state.
Arrows at the red trajectory show the movement direction of the
corresponding Majorana point.
} \label{fig:ghz}
\end{figure}

 Eq.~(\ref{eq:e1}) shows that the dynamics of the Majorana points for such a family is relatively simple. A single star representing  $|\phi_1\rangle$ stays at the north pole all the time,  while the second star makes a loop around the Bloch sphere. In particular for $\beta=0$ or $\pi$ we have both points at north pole and for $\beta=\pi/2$ we have the Bell state $\frac{1}{\sqrt{2}}(|01\rangle+|10\rangle)$ with one point at the north and the second at the south pole. As shown in Fig.~\ref{fig:ve1} the velocity of this second Majorana point is highly nonlinear  In the case considered the velocity is anticorrelated with the amount of entanglement. For the maximally entangled Bell state which emerges at $\beta=\pi/2$ the velocity reaches its minimum while the maximum is attained for the separable state $|00\rangle$.

 In general, the dynamics of the stars defining the stellar representation of a state is not as simple. For instance, for a family $\frac{1}{\sqrt{K(\beta)}}\big(\cos\beta\mid000\rangle+\sin\beta\mid GHZ\rangle\big)$, with $K(\beta)$ a suitable normalization factor, one obtains the motion of $3$ points shown in Fig.~\ref{fig:ghz}.

\subsection{Permutation invariant Hamiltonians}
Consider a more general dynamics defined by a one-parameter subgroup
 of the unitary group,
\begin{equation}
U=e^{-i\beta H},
\label{eq:u}
\end{equation}
where $H$ is a Hermitian Hamiltonian.
Of interest are only transformations preserving the permutation symmetry of
states. We are going to study an exemplary family of two-qubit Hamiltonians,
parametrized by  indices $i,j=0,\dots 3$ with $i \ne j$,
\begin{equation}
H_{ij}=\frac{1}{2} (\sigma_i\otimes\sigma_j+\sigma_j\otimes\sigma_i),
\label{eq:h}
\end{equation}
where $\sigma_i$ represents the Pauli matrix and $\sigma_0={\mathbb{I}}$.
Using $\sigma_i\sigma_j=i\epsilon_{ijk}\sigma_k+\mathbb{I}\delta_{ij}$
we obtain,
\begin{eqnarray}\nonumber
&U=-\frac{1}{2}\Big(\cos\beta(\mathbb{I}\otimes\mathbb{I}+\sigma_k\otimes\sigma_k)+&\\
&i\sin\beta (\sigma_i\otimes\sigma_j+\sigma_j\otimes\sigma_i)+(\mathbb{I}\otimes\mathbb{I}-\sigma_k\otimes\sigma_k)\Big).&
\end{eqnarray}

Clearly $U$ preserves the permutation symmetry as it acts on both particles
in the same way. This is also true for any linear combination
 of generators $H_{ij}$ of the form (\ref{eq:h}).
Taking thus
$$H=\frac{H_{23}+H_{03}}{\sqrt{2}}=-\frac{1}{2\sqrt{2}}
\Big(\sigma_2\otimes(\sigma_0+\sigma_3)+(\sigma_0+\sigma_3)\otimes\sigma_2\Big)$$
we obtain:
\begin{equation}
  U=\left(\begin{array}{cccc}
 \cos\beta & -\frac{\sin\beta}{\sqrt{2}} & -\frac{\sin\beta}{\sqrt{2}} & 0 \\
 \frac{\sin\beta}{\sqrt{2}} & \frac{1}{2} (1+\cos\beta) & \frac{1}{2} (-1+\cos\beta) & 0 \\
 \frac{\sin\beta}{\sqrt{2}} & \frac{1}{2} (-1+\cos\beta) & \frac{1}{2} (1+\cos\beta) & 0 \\
 0 & 0 & 0 & 1
\end{array}
\right)
\end{equation}
which transforms the state $|00\rangle$ into $|\psi_1\rangle=\cos\beta
|00\rangle+\frac{\sin\beta}{\sqrt{2}}(|01\rangle+|10\rangle)$ as in the
example considered previously.
Note that the operator $U$ which generates the same dynamics of stars differs from  the
operators $U_1$ and $U_2$ defined in (\ref{eq:u12}).

\subsection{Unitary dynamics of symmetric states of $(N-1)$
qubits and corresponding dynamics of a single quNit}

As discussed in previous sections, any symmetric pure state of $n$--qubit system
can be associated with a state of a simple system of size $N=n+1$.
Thus any unitary dynamics of the composite system, which
preserves the permutations symmetry, induces a certain dynamics
in ${\cal H}_N$. To make such a link explicit consider
a permutation symmetric state  $\psi\in{\cal H}_{2}^{\otimes n}$
which can be represented in the standard basis,
$\{|0\rangle^{\otimes k}\otimes|1\rangle^{\otimes(n-k)}\}_{0\leq k\leq n}$.
Define a a unitary transition matrix $T$ that transforms this basis into an orthonormal
basis with first $n+1$ permutation symmetric vectors and remaining $2^n-(n+1)$ vectors,
which span a basis in the orthogonal subspace.
Denote $|\psi '\rangle=T^{-1}|\psi\rangle
={\phi \choose 0}$ where
$|\phi\rangle \in{\cal H}_{n+1}$, $|0\rangle \in{\cal H}_{2^n-n-1}$
and  take any  unitary operator $U$, acting on ${\cal H}_{2}^{\otimes n}$,
which preserves the permutational symmetry.
We have then:
\begin{eqnarray} \nonumber
|\tilde{\psi}\rangle &=&U |\psi\rangle , \\
T^{-1}|\tilde{\psi}\rangle &=&T^{-1}UTT^{-1}|\psi\rangle,  \\ \nonumber
|\tilde{\psi}'\rangle &=&U'|\psi'\rangle,
\label{eq:uprim}
\end{eqnarray}
where $U'=T^{-1}UT$ and $|\tilde{\psi}'\rangle=T^{-1}|\tilde{\psi}\rangle$. As $U$ preserves the permutation symmetry
 condition  Eq.~(\ref{eq:uprim})
implies the following block structure of the matrix $U'$:

\begin{equation}
U'=\left(\begin{array}{c|c}
V & 0 \\ \hline
0 & W
      \end{array}\right).
\label{eq:block}
\end{equation}
Here $V$ and $W$ denote unitary matrices of size
$n+1$ and $2^n-(n+1)$, respectively.
The matrix $V$ acts on the $(n+1)$--dimensional subspace of permutation symmetric vectors.
These vectors are in one to one correspondence with vectors spanning an orthonormal basis
in ${\cal H}_{n+1}$ -- see e.g. \cite{BR45}.
In other words the matrix $U$ representing a permutation symmetry preserving dynamics
has to be reducible.
The unitary block $V$ of the rotated matrix, $U'=T^{-1}UT$,
defines thus a unitary dynamics of the one quNit system,
associated with the $n$-qubit dynamics $U$, which preserves the permutational symmetry.

\subsection{A three--qubit example}
To illustrate the above reasoning let us analyze a $3$--qubit example.
We use the natural basis,
$\big(|000\rangle,|001\rangle,|010\rangle,|011\rangle,|100\rangle,|101\rangle,|110\rangle,|111\rangle \big)$
and the unitary transition matrix $T$ reads:
\begin{equation}
 \small{T=\left(
\begin{array}{cccccccc}
 1 & 0 & 0 & 0 & 0 & 0 & 0 & 0 \\
 0 & \frac{1}{\sqrt{3}} & 0 & 0 & \frac{1}{\sqrt{2}} & \frac{1}{\sqrt{6}} & 0 & 0 \\
 0 & \frac{1}{\sqrt{3}} & 0 & 0 & -\frac{1}{\sqrt{2}} & \frac{1}{\sqrt{6}} & 0 & 0 \\
 0 & 0 & \frac{1}{\sqrt{3}} & 0 & 0 & 0 & \frac{1}{\sqrt{2}} & \frac{1}{\sqrt{6}} \\
 0 & \frac{1}{\sqrt{3}} & 0 & 0 & 0 & -\sqrt{\frac{2}{3}} & 0 & 0 \\
 0 & 0 & \frac{1}{\sqrt{3}} & 0 & 0 & 0 & -\frac{1}{\sqrt{2}} & \frac{1}{\sqrt{6}} \\
 0 & 0 & \frac{1}{\sqrt{3}} & 0 & 0 & 0 & 0 & -\sqrt{\frac{2}{3}} \\
 0 & 0 & 0 & 1 & 0 & 0 & 0 & 0
\end{array}
\right)}.
\end{equation}
Using formula (\ref{eq:u}) we construct permutation symmetry preserve operator $U$
taking $H=\sum\limits_\pi \sigma_1\otimes\sigma_3\otimes\left(\begin{array}{cc}
1 & 0 \\
0 & 0   \end{array}\right)$,
where $\sum\limits_\pi$ stands for the sum over all permutations.
The matrix $U$ obtained in this way does not have any special structure.
However, the transformation $T$ brings it to the matrix $U'=T^{-1}UT$
which enjoys the block structure as in (\ref{eq:block}),
$$\tiny{V=\left(\begin{array}{cccc}
 \frac{1}{4} (1+3 cos(4 \beta)) & -\frac{1}{2} i \sqrt{3} sin(4 \beta) & 2 \sqrt{3} cos(\beta)^2 sin(\beta)^2 & 0 \\
 -\frac{1}{2} i \sqrt{3} sin(4 \beta) & cos(4 \beta) & \frac{1}{2} i sin(4 \beta) & 0 \\
 2 \sqrt{3} cos(\beta)^2 sin(\beta)^2 & \frac{1}{2} i sin(4 \beta) & \frac{1}{4} (3+cos(4 \beta)) & 0 \\
 0 & 0 & 0 & 1
\end{array}\right)}.$$

The matrix $V$ determines thus the corresponding dynamics of a simple system of size
$n+1=4$, while the remaining part,
$$\tiny{W=\left(\begin{array}{cccc}
 cos(\beta) & 0 & -\frac{1}{2} i sin(\beta) & \frac{1}{2} i \sqrt{3} sin(\beta) \\
 0 & cos(\beta) & \frac{1}{2} i \sqrt{3} sin(\beta) & \frac{1}{2} i sin(\beta) \\
 -\frac{1}{2} i sin(\beta) & \frac{1}{2} i \sqrt{3} sin(\beta) & cos(\beta) & 0 \\
 \frac{1}{2} i \sqrt{3} sin(\beta) & \frac{1}{2} i sin(\beta) & 0 & cos(\beta)
\end{array}\right)}.$$
is irrelevant for the corresponding dynamics in ${\cal H}_{n+1}$.

\subsection{Dynamics for two qubit case}

From (\ref{eq:u}) we obtain a differential equations for the dynamics
\begin{equation}
i\partial_\beta\mid\psi\rangle=H\mid\psi\rangle.
\label{eq:row}
\end{equation}
which can be also written in terms of the angles $\theta_i$ and the phases
$\Phi_i$, $i=1,2$ characterizing the positions of the Majorana points.
 In the general case such equations are not easy to solve.
Therefore we will restrict here our attention to the special case of $n=2$ qubits.

\begin{figure}
\includegraphics[width=5.5cm]{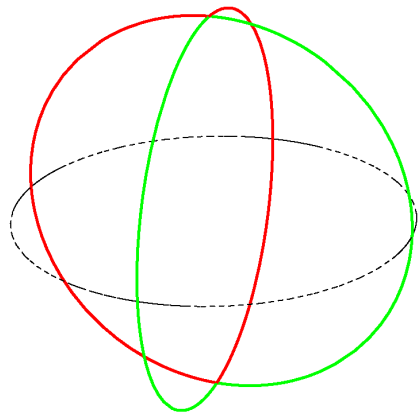}
\caption{(color online) Trajectories of MP for the state
$|\psi_2\rangle=\cos\beta\mid00\rangle-\sin\beta\mid11\rangle$. Red and green
line represent trajectories of two points for $\beta\in[0,\pi]$, black dashed line represents the
equator. Both stars travel from the north to the south pole,
 where they meet at $\beta=\pi/2$.
Later on they return back to the north pole.}
\label{fig:2q-traj}
\end{figure}

Let us take $H=-\sigma_1\otimes\sigma_2-\sigma_2\otimes\sigma_1$. Using
Eq.~(\ref{eq:u}) we obtain the operator $U$ which generates dynamics of the
MPs of state $|\psi_2\rangle=\cos\beta\mid00\rangle-\sin\beta\mid11\rangle$.
The corresponding dynamics of the stars on the sphere
with respect to the time $\beta$ is shown in Fig.~\ref{fig:2q-traj}.

For simplicity we restrict our analysis to $\beta\in[0,\pi/2]$, where
$\theta_1=\theta_2$, $\Phi_1=\Phi_2+\pi$ and the phases $\Phi_1$
and $\Phi_2$ do not change. Using Eq.~(\ref{eq:row}) we obtain a differential equation for the
polar angle $\theta=\theta_1=\theta_2$,
\begin{equation}
  \dot{\theta}=\frac{3+\cos 3\theta}{2\sin\theta}.
\label{eq:me}
\end{equation}

\begin{figure}
\includegraphics[width=6cm]{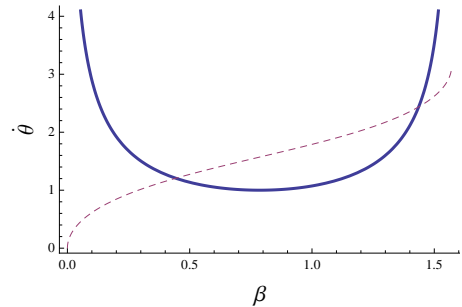}
\caption{(color online) Velocity $\dot{\theta}$ for the state $|\psi_2\rangle=\cos\beta\mid00\rangle-\sin\beta\mid11\rangle$. Full line is $\dot{\theta}$ in the function of $\beta$, dashed line is $\theta(\beta)$.}
\label{fig:2q-v}
\end{figure}

As shown in Fig.~\ref{fig:2q-v} the velocity diverges when $\theta$ approaches
to zero or $\pi$ and both stars are located in a single point.
The velocity $\dot{\theta}$ reaches its minimum for a Bell state obtained
for $\theta=\pi/2$

\begin{figure}
\includegraphics[width=6cm]{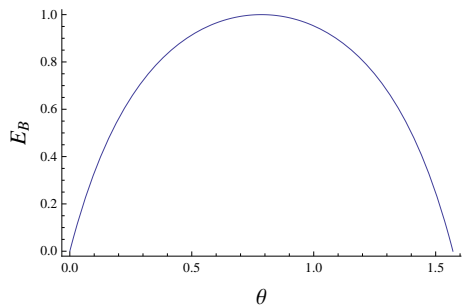}
\caption{Dependence of the barycentric entanglement measure $E_B$ on the
phase $\beta$ for the state
$|\psi_2\rangle=\cos\beta\mid00\rangle-\sin\beta\mid11\rangle$.}
\label{fig:2q-r}
\end{figure}

It is instructive to compare the velocity of stars during the unitary dynamics of a state
parametrized by the the phase $\beta$
with changes of the barycentric measure $E_B$ -- see Fig.~\ref{fig:2q-r}.
Observe an anticorrelation between entanglement
measure and $\dot{\theta}$ - the maximal velocity occurs for the minimal
entanglement and \textit{vice versa}.
This observation implies that product states tend to gain
entanglement during a relatively small interaction time.

\section{Conclusions}

In this work we presented a homogeneous approach to study the structure of pure quantum states describing two physical problems: a simple system consisting of $n+1$ levels and the class of states of an $n$--qubit system, symmetric with respect to  permutations of all subsystems. Making use of the Majorana--Penrose representation one can find a direct link between these two cases as any constellation of $n$ stars on the sphere determines a quantum state in both setups. In particular, product states of the multi--qubit system, for which all Majorana points coalesce in a single point, correspond to spin coherent states of the simple system. To find a direct relation between two problems one can identify  the orthogonal basis of the eigenstates of the angular momentum operator $J_z$ acting on ${\cal H}_{n+1}$ with the set of $(n+1)$ Dicke states which span the complete basis in the subspace of symmetric states of the composite system.

Physical properties of a given pure state can be thus related to the distribution of the corresponding  collection of $n$ points on the sphere. The variance of this distribution, related to the radius of the barycenter inside the ball, can be thus used to characterize the degree of non spin coherence of the states of a simple system or the degree of entanglement for the composite systems. The proposed barycentric measure of quantum entanglement achieves its maximum for these states, for which the barycenter of the corresponding Majorana points is located at the center of the ball. This class of states includes the Bell state of a two-qubit system, the GHZ state of a three qubit system and several states distinguished by being most distant from the set of separable states and called 'Queen of Quantum' \cite{GBB10}. In the case of four--qubit states we have explicitly described a two-parameter class of extremal symmetric states for which the barycentric measure achieves the maximal value, $E_B=1$.
All these states are also highly entangled with respect to the geometric
measure $E_G$, which for them belongs to the interval $[1, \log_2 3]$.

It is convenient to define the composition of
two permutation symmetric states of an arbitrary number $n$ and $m$ qubits,
each of them described by $n$ and $m$ Majorana points, respectively.
The composition is characterized by the collection of $n+m$ stars on the sphere
and represents a symmetric state of $m+n$ qubits. Making use of these notion
we show that the state $|GHZ_4\rangle$ can be interpreted
as a composition of two two--qubit Bell states
and construct a family of maximally entangled multiqubit
pure states with $E_B=1$.

Any unitary dynamics acting on the $n$--qubit system is described by a matrix $U$ of order $2^n$. Under assumption that the dynamics does not break the permutation symmetry, the matrix $U$ is reducible and can be written as a direct sum of two unitary matrices, $U=V \oplus W$. The matrix $V$ or order $n+1$ describes the unitary dynamics in the subspace of symmetric states of the composed system or the corresponding unitary dynamics in the space of all states of the $(n+1)$--level system. A unitary dynamics of a quantum  pure state leads to a non-linear dynamics of the corresponding stars of its Majorana representation. In a simple model evolution investigated for a two qubit system the velocity of stars is small if they are far apart, what corresponds to the highly entangled states, and it increases as the stars get together  and the state is close to be separable.

\medskip

It is a pleasure to thank D.~Braun, P.~Braun, D.~Brody,
D.~Chru{\'s}ci{\'n}ski and  O.~Giraud
for fruitful discussions and helpfull correspondence.
Financial support by the grant number N N202 261938 of
 Polish Ministry of Science and Higher Education is gratefully acknowledged.

\end{document}